\documentclass[twocolumn,trackchanges]{aastex7}

\usepackage{subfig}
\usepackage{xcolor}

\begin{document}

\title{Flare genesis in relativistic jet: Disentangling the drivers of variability in the blazar 4C +27.50}

\author[orcid=0009-0008-3960-1213]{Joysankar Majumdar}
\affiliation{Department of Physics, Institute of Science, Banaras Hindu University, Varanasi - 221005, India}
\affiliation{Astronomical Observatory, University of Warsaw, Al. Ujazdowskie 4, 00-478 Warsaw, Poland}
\email{joysankarmajumdar@gmail.com}

\author[orcid=0009-0007-1587-2871]{Sakshi Maurya} 
\affiliation{Department of Physics, Institute of Science, Banaras Hindu University, Varanasi - 221005, India}
\affiliation{Astronomical Observatory, University of Warsaw, Al. Ujazdowskie 4, 00-478 Warsaw, Poland}
\email{mauryasakshi158@gmail.com}

\author[orcid=0000-0002-1173-7310]{Raj Prince} 
\affiliation{Department of Physics, Institute of Science, Banaras Hindu University, Varanasi - 221005, India}

\email[show]{priraj@bhu.ac.in}
\correspondingauthor{Raj Prince}

\begin{abstract}
Recently, blazar 4C +27.50 was found to be flaring in gamma-rays since its detection with Fermi-LAT in 2008.
For the first time, a dedicated temporal and spectral study of the blazar 4C +27.50 has been performed in this work to understand the nature of this object. We used the Bayesian block algorithm to identify four flaring states and one quiet state in the 4-year-long Fermi-LAT light curve. Simultaneous broadband flaring episodes have been observed, and a significant correlation is seen between optical and $\gamma$-ray emission, suggesting the co-spatial origin of the broadband emission. The variation of fractional variability amplitude with respect to frequency shows a nearly double hump structure similar to broadband SED. The fastest flux doubling time in the 1-day binned $\gamma$-ray light curve is found to be about 7.8 hours. A curvature in $\gamma$-ray spectra has been observed, possibly caused by a stochastic particle acceleration process rather than radiative cooling. No evident correlation was found in the $\gamma$-ray flux-index plot, but a clear harder-when-brighter trend is observed in the X-ray flux-index plot. A one-zone leptonic model has been implemented to understand broadband emission during the quiet and flaring states, and the variation of the jet parameters has been investigated. Broadband SED modeling suggested that an enhancement in the magnetic field and particle energy density, together with a harder particle spectrum, might have caused the flaring events, while the bulk Lorentz factor shows no systematic change.

\end{abstract}


\section{Introduction} \label{sec:intro}
Blazar is a distinctive class of active galactic nucleus (AGN), generally described by the infall of matter into a supermassive black hole (SMBH) and the emergence of oppositely directed sub-parsec to megaparsec jets, with one jet aligned along the observer's line of sight \citep{1995PASP..107..803U}. It is observed that blazars are the most powerful and stable emitters of non-thermal radiation in the universe \citep{2019ARA&A..57..467B}. Blazars are further categorized into flat-spectrum radio quasars (FSRQs), which possess broad emission lines, and BL Lac objects, which exhibit weak or completely absent emission or absorption lines \citep{2007ApJ...662..182P}. The radiation emitted from the blazar's jet covers the electromagnetic spectrum, ranging from radio to $\gamma$-ray \citep{1978PhyS...17..265B}. Abrupt flaring episodes are observed in the multi-wavelength emissions with wide-ranging timescales from minutes to years \citep{2020NatCo..11.4176S,2021MNRAS.502.5245P}.

Two evident humps are observed in the broad-band spectral energy distribution (SED) of blazars, which is ascribed to radiative losses experienced by a non-thermal electron distribution \citep{2010ApJ...716...30A}. The low-energy hump is from synchrotron radiation emitted by a population of relativistic electrons in the jet as they lose energy in a magnetic field. The high-energy hump is from the inverse Compton (IC) process, where the photons involved in the IC process may come from internal to the jet (synchrotron self-Compton; \citealt{2009ApJ...704...38S}) or external radiation fields (external Compton; \citealt{1992A&A...256L..27D,1994ApJ...421..153S}) and hadronic radiative processes including Photo-meson interaction, Bethe-Heitler pair production and Proton synchrotron radiation \citep{2020Galax...8...72C,2025JHEAp..4800402M}. Thus, conducting timing and SED studies of blazars can help uncover valuable insights about the processes occurring at the onset of the jet.

It is very complex to model the SED of a blazar. As shown by \citet{2018A&A...619A.159M}, SED modeling of FSRQ PKS 1510-089 during low states between 2012 and 2017 suggested that the $\gamma$-ray emission region is just outside the BLR. \citet{2021MNRAS.502.5245P} showed that the emission region of FSRQ OQ 334 during the flaring states of 2018 and 2019 was just inside the BLR using SED modeling. Blazar 4FGL J1520.8–0348 is classified as FSRQ, and SED modeling suggested that the emission region is outside the BLR and inside the dusty torus \citep{2024ApJ...965..112R}. Thus, SED modeling gives insightful information about the reason behind the flares in blazars.

4C +27.50 is a high distant AGN with a redshift (z) of 1.2551 \citep{2022ApJS..260...33S}. Earlier, 4C +27.50 is classified as Quasar by \citet{1981A&AS...45..367K} and later it is classified as FSRQ \citep{2018ApJ...866..137L,2022ApJ...925...40X,2022ApJ...925...97P}. It has been in the low $\gamma$-ray flux state since 2008. It has been observed flaring in $\gamma$-ray by Fermi-LAT in 2022 as reported by Atel \#15549 \citep{2022ATel15549....1L}. After this flaring episode, the source went to a quiescent period, and in 2024 it again flared in $\gamma$-ray as reported by Atel \#16718 \citep{2024ATel16718....1C} and simultaneously flaring events in optical and NIR bands are also reported by Atel \#16724 \citep{2024ATel16724....1B} and Atel \#16863 \citep{2024ATel16863....1C}. Recently, at the beginning of 2025, it again started flaring in $\gamma$-ray as reported by Atel \#17035 \citep{2025ATel17035....1B}. Source is interesting in the sense that it is located at a relatively high redshift and has shown a strong broadband flare, which is unique compared to other high redshift objects. In literature, high redshift blazars have been studied \cite{2020MNRAS.498.2594S}, but their flares are not as strong as this source; also, in most cases, they do not show broadband flares at high redshift.
Thus, in this study, we performed a comprehensive timing and spectral analysis of 4C +27.50 during its flaring and quiescent periods to understand the reason behind this dynamics.

In section 2, we discussed the analysis procedure of various telescopes, followed by the results and discussions in section 3. In section 4, we summarized our findings.

\begin{figure*}
    \centering
    \includegraphics[width=0.9\linewidth]{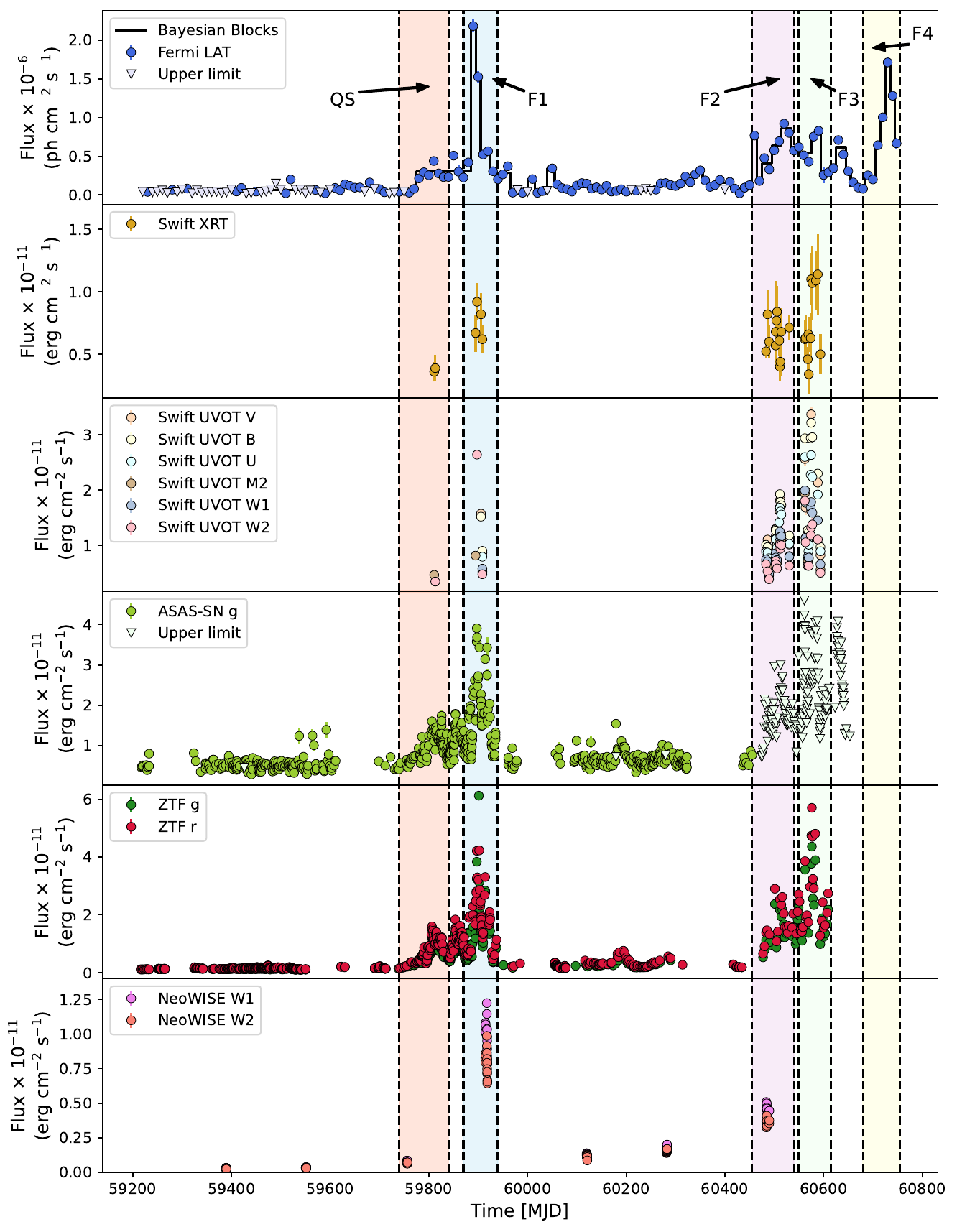}
    \caption{Multi-wavelength light curve of 4C +27.50 from MJD 59215 to 60751.}
    \label{fig:all_lcs}
\end{figure*}

\section{Multi-wavelength Observations and Data Analysis} \label{sec:data_analysis}

\subsection{Fermi-LAT}
NASA's Fermi satellite, launched in 2008, features the Large Area Telescope (LAT) as its main instrument for detecting gamma rays \citep{2009ApJ...697.1071A}. LAT operates in the energy range of 20 MeV to over 300 GeV with a peak effective area of $\sim$8000 cm$^2$ and a field of view of approximately 2.4 steradians.

We analyzed collected LAT data\footnote{\url{https://fermi.gsfc.nasa.gov/cgi-bin/ssc/LAT/LATDataQuery.cgi}} of 4C +27.50 (4FGL J2321.9+2734) from 2021 (MJD 59215) to 2025 (MJD 60751), including multiple flares. We adopted \texttt{Fermipy} \citep{2017ICRC...35..824W} to analyze LAT data. Our analysis was conducted within the energy range of 100 MeV to 500 GeV, using \texttt{evclass} = 128, \texttt{evtype} = 3 parameters and filter = \texttt{DATA\_QUAL$>$0 \&\& LAT\_CONFIG==1}, while applying a zenith angle cut of $>$ 90$^\circ$ to reduce contamination from Earth’s limb $\gamma$-rays. We selected 15$^\circ$ centered on the source
position as our region of interest (ROI) to build the model along with the \texttt{gll\_iem\_v07} Galactic diffuse emission model and the \texttt{iso\_P8R3\_SOURCE\_V3\_v1} isotropic background model\footnote{\url{https://fermi.gsfc.nasa.gov/ssc/data/access/lat/BackgroundModels.html}}.

The spectral models and parameters for sources within the ROI were obtained from the fourth Fermi source catalog (4FGL-DR4; \citealt{2023arXiv230712546B}). We optimized all sources using \texttt{Fermipy}, and then all the parameters of 4C +27.50, the Galactic diffuse emission model, and the isotropic background model were kept free. The norm of all the sources within the 3$^\circ$ centered on the source was also kept free. We have localized 4C +27.50 based on the Test Statistics (TS) map using the \texttt{localize()} function of \texttt{Fermipy}. Then, we performed the fitting and generated the light curve and SED with 4 energy bins per decade.

\subsection{Swift-XRT} \label{subsec:XRT}
We used the Swift-XRT data products generator\footnote{\url{https://www.swift.ac.uk/user_objects/index.php}} to generate the spectrum \citep{2009MNRAS.397.1177E}, which used \texttt{HEASOFT v6.32} at the time we produced the light curves for all the Swift observations from MJD 59215 to MJD 60751 in the energy range 2 - 10 keV. It also estimates the flux corresponding to each observation, and we used these fluxes to produce the X-ray light curve of 4C +27.50.

For the SED of particular epochs, combined spectra were produced from the Swift-XRT data products generator with a simple power law model. The non-Galactic absorption component was calculated based on the redshift value of 1.255 \citep{2022ApJS..260...33S}. The processed spectra files were grouped using the task \texttt{grppha} and again modeled with \texttt{xspec}. Using \texttt{iplot}, we saved the spectra for use in broadband SED.

\subsection{Swift-UVOT} \label{subsec:UVOT}
The Swift-UVOT instrument observed 4C +27.50 using three optical filters (V, B, and U) and three ultraviolet filters (W1, M2, and W2). We retrieved the UVOT data from the HEASARC Data Archive and analyzed it using \texttt{HEASoft} version 6.34 with \texttt{CALDB} version 20240201. For each filter, we combined the observations over the specified period using the \texttt{UVOTIMSUM} task and determined the target's magnitude with \texttt{UVOTSOURCE}. The magnitudes were corrected for Galactic extinction following \citet{2011ApJ...737..103S}. The source magnitude was extracted using a 4.5 arcsec circular region centered on the source, while the background magnitude was obtained from a 10 arcsec circular region nearby. Finally, we converted the corrected magnitudes to fluxes using the zero points from \citet{2011AIPC.1358..373B} and the conversion factors from \citet{2016MNRAS.461.3047L}.

\subsection{Archival data}
We accessed the archival Zwicky Transient Facility (ZTF; \citealt{2019PASP..131a8002B}) optical data of g and r bands from NASA/IPAC Infrared Science Archive\footnote{\url{https://irsa.ipac.caltech.edu/cgi-bin/Gator/nph-dd}} within a cone of 5 arcsec by keeping the source at the center of it. We have also accessed the All-Sky Automated Survey for Supernovae (ASAS-SN) optical data using ASAS-SN Sky Patrol V2.0\footnote{\url{http://asas-sn.ifa.hawaii.edu/skypatrol/}} \citep{2023arXiv230403791H} in g band only. The optical data from ZTF and ASAS-SN were corrected for Galactic extinction following the open source code of the \texttt{VOU-Blazars} tool \citep{2020A&C....3000350C}.

We have accessed the IR archival data from NeoWISE \citep{2011ApJ...731...53M}, which is the asteroid-hunting portion of Wide-field Infrared Survey Explorer (WISE) through NASA/IPAC Infrared Science Archive\footnote{\url{https://irsa.ipac.caltech.edu/cgi-bin/Gator/nph-dd}}.

\section{Results and Discussions} \label{sec:results}
\subsection{Flare selection}
We adopted the Bayesian blocks (BB) algorithm \citep{2013ApJ...764..167S} to detect flares in the $\gamma$-ray light curve. We used a PYTHON package \texttt{lightcurves} to implement the BB algorithm with a false-alarm probability (p$_{0}$) of 0.05 in the $\gamma$-ray light curve, and it is shown in the first panel of Figure \ref{fig:all_lcs} with a black line. When the $F_{BB}$ $>$ 3$\bar{F}$, then it is considered as a flare, where $F_{BB}$ is the flux computed using the BB algorithm and $\bar{F}$ is the average flux. Using this BB algorithm, we selected the first flare (F1) from MJD 59870 to MJD 59940. We considered the MJD 60455 to MJD 60540 time range as the second flare (F2) and the MJD 60550 to MJD 60615 time range as the third flare (F3) because in this time range, X-ray, UV, and optical showed a huge flux enhancement, and the MJD 60680 to MJD 60755 time range as the fourth flare (F4). We did not consider F4 for a multi-wavelength study as data from other wavelengths is unavailable. We submitted the target of opportunity request for SWIFT observations, but observations were not possible due to the Sun constraint. We considered the period from MJD 59740 to MJD 59840 as the quiescent state (QS). Thus, QS, F1, F2, F3 and F4 are marked in Figure \ref{fig:all_lcs}.

\subsection{Multi-wavelength light curves}
We showed the multi-wavelength light curve for 4C +27.50 from MJD 59215 to MJD 60751 in Figure \ref{fig:all_lcs}. The first panel shows the Fermi-LAT light curve in the energy range of 100 MeV to 500 GeV with 10-day binning. The second and third panels show the Swift-XRT and Swift-UVOT light curves, respectively. Light curves from the ASAS-SN g-band, ZTF-g, and ZTF-r are shown in the fourth and fifth panels, respectively. The last panel shows the NeoWISE light curve. The observed light curves across the waveband show strong variability. Since the gamma-ray light curve is well sampled, we used it to identify the flaring state using the Bayesian-block methodology as mentioned in the above section. We identified the four bright flaring states and named them as F1, F2, F3, and F4. Since there are no other bands' observations available for F4, so, we focused only on the first three flares. Flare F1 happens to be the brightest among all and the first flare ever observed from this source. The division of F2 and F3 is also based on different flux states in XRT and UVOT. F3 is observed to be the brightest in the X-ray and optical-UV regime. 
The source was observed to be the brightest in the infrared (IR) during state F1 compared to F2. During F3, we do not have any IR observations. Based on the broadband light curves, we can say that there is a one-to-one correlation between gamma-ray, X-ray, UV-optical, and IR. Later, this correlation is confirmed with a cross-correlation study. The simultaneous flaring across the waveband is seen earlier as well in many blazars, suggesting they are produced simultaneously at the same location and can be explained under the one-zone emission model.

\subsection{Variability study} \label{subsec:var_study}
\begin{figure}[ht]
    \centering
    \includegraphics[width=1\linewidth]{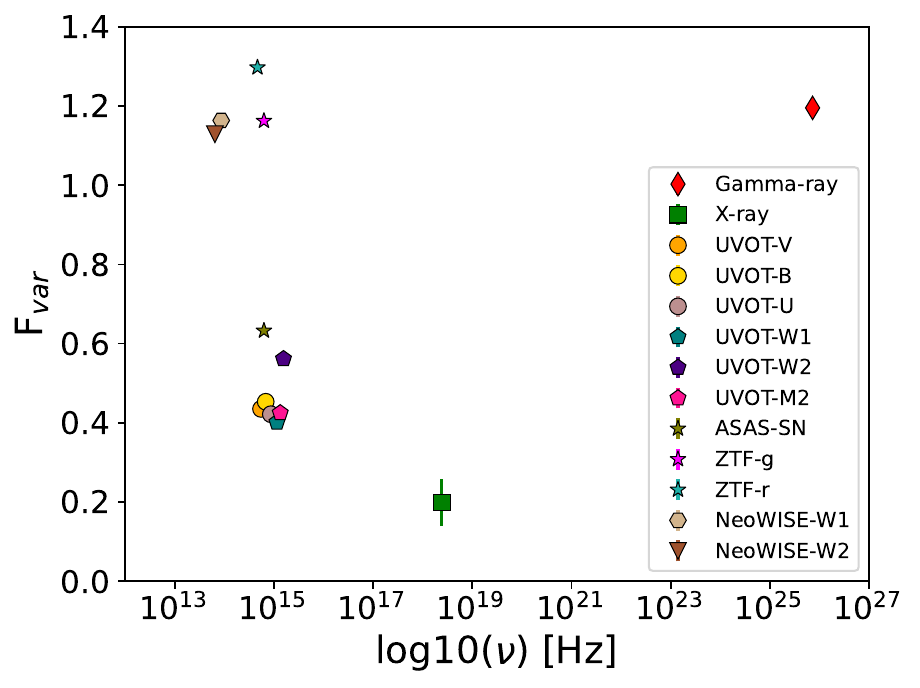}
    \caption{Fractional variability of 4C +27.50 in different wavebands.}
    \label{fig:frac_var}
\end{figure}

Variability in blazars is a random phenomenon that occurs at all wavelengths and timescales. It is more noticeable during the flare, and the flare profile is influenced by the energy dissipation and particle acceleration in the jet. The amplitude of variation is influenced by the jet parameters like magnetic field, viewing angle, particle density, and the efficiency of the acceleration process \citep{2018A&A...617A..59K}. \citet{1987ApJ...323..516E} introduced the fractional root mean square (rms) variability parameter (F$_{var}$) to determine the variability amplitude. Fractional variability can be computed using the relation provided in \citet{2003MNRAS.345.1271V} and is used to compare the variability amplitudes across the multi-wavebands,

\begin{equation}
    F_{var} = \sqrt{\frac{S^2-\sigma^2}{r^2}}
\end{equation}

\begin{equation}
    err(F_{var}) = \sqrt{  \left(\sqrt{\frac{1}{2N}} \frac{\sigma^2}{r^2 F_{var}}\right)^2  +  \left(\sqrt{\frac{\sigma^2}{N}} \frac{1}{r}\right)^2 }
\end{equation}
where $S^2$ is the sample variance, $\sigma^2$ is the mean square uncertainty of each observation, $N$ is the number of sample points and $r$ is the sample mean. The fractional variabilities for all wave bands are tabulated in Table \ref{tab:f_var} and plotted in Figure \ref{fig:frac_var}. However, the fractional variability also depends upon the binning of the data; it is clear that the source is the most variable in optical (ZTF r) and then $\gamma$-ray, followed by IR, and X-ray.

The source variability can also be quantified using the flux doubling/halving timescale ($t_d$), which represents the time taken for the flux to increase or decrease by a factor of two between successive intervals, given by

\begin{equation}
    F(t_2) = F(t_1)\,2^{\frac{t_2-t_1}{t_d}}
\end{equation}
where F($t_1$) and F($t_2$) are the fluxes measured at time $t_1$ and $t_2$. We produced a 1-day binned $\gamma$-ray light curve during F1 and found the minimum flux doubling/halving timescale ($t_d$) to be $\sim$7.8 hours.

This timescale can be used to constrain the size of the emission region, assuming it is a homogeneous spherical blob, using the causality argument. We estimated the radius of the spherical blob (R) using the relation \citep{1995MNRAS.273..583D, Dermer_2009}
\begin{equation}
    R \leq c\,{t_{d}}\,\delta\,(1+z)^{-1}
\end{equation}
where $z=1.2551$ \citep{2022ApJS..260...33S} and $\delta$ is the Doppler factor. We initially adopted $\delta = 23.9^{+21.09}_{-12.12}$, as estimated by \citet{2018ApJ...866..137L}, for the radio-emitting jet component. The $\delta$ was derived through modeling of a 15 GHz light curve using a Bayesian hierarchical code, which decomposes variability into flaring events and infers the variability brightness temperature. By comparing this with the equipartition brightness temperature, they estimated the variability Doppler factor. However, this $\delta$ specifically characterizes the radio core, and the broadband emission modeled in this work may originate from distinct, possibly more compact regions with different $\delta$. Despite this, we consider $\delta = 23.9$ a reasonable first-order approximation, providing a consistent framework for our modeling. For $t_d = 7.8$ h we found $R \leq 8.9 \times 10^{15}$ cm.

The high-energy emission from blazars is explained by internal shocks \citep{2013EPJWC..6102007R} or magnetic reconnection \citep{2015MNRAS.450..183S}. Under the internal shock model or colliding shell model, the turbulence in the jet can cause two magnetized plasma layers with different velocities to collide in the emission region, producing the variability. The following estimate applies only to this colliding-shell scenario. The first layer goes with velocity $v_1$ and the second layer goes with velocity $v_2$. Now we assume, $v_2>v_1$ and $v_2 \simeq c$. The delay between the ejections of the two layers appears as variability to the observer. The first layer is ejected at time zero and the second after a delay $t_d'$. The collision occurs at a time $t$ after the ejection of the second layer, when the distance of the first layer, $R_1 = v_1(t+t_d')$, equals the distance of the second layer, $R_2 = v_2t$, and it implies $t=\frac{v_1}{v_2-v_1}t_d'$. Here $t_d$ is the observed flux doubling time and $t_d' = t_d/(1+z)$ is the corresponding delay in the rest frame of the source.
 The distance of the collision, which essentially restricts the emission region location, is
\begin{equation}
    R_H=R_2=v_2t=\frac{v_1v_2}{v_2-v_1}t_d'=\frac{v_1}{1-\frac{v_1}{v_2}}t_d'
\end{equation}
Now multiplying $1+\frac{v_1}{v_2}$ to numerator and denominator,
\begin{equation}
    R_H=\frac{v_1(1+\frac{v_1}{v_2})}{1-(\frac{v_1}{v_2})^2}\,t_d'
\end{equation}
Following the assumption of $v_2>v_1$ and $v_2 \simeq c$, the Lorentz factor $\Gamma=\frac{1}{\sqrt{1-v_1^2/c^2}}$ and now the emission region location becomes
\begin{equation}
    R_H \simeq \frac{2v_1}{1-(\frac{v_1}{v_2})^2}t_d'=2v_1\Gamma^2t_d'
\end{equation}
and for the ultra-relativistic case \citep{2019Galax...7...28R},
\begin{equation}
    R_H \simeq 2c\Gamma^2t_d' = 2c\Gamma^2\frac{t_d}{1+z}
    \label{eq:rh}
\end{equation}
We do not assume $\delta = \Gamma$. For the viewing angle $\theta = 0.5^\circ$ adopted in the SED modeling (Section~\ref{subsec:sed}), $\delta = [\Gamma(1-\sqrt{1-\Gamma^{-2}}\cos\theta)]^{-1} = 23.9$ gives $\Gamma \approx 12$, and the shortest variability time, $t_d = 7.8$ hrs, gives $R_H \approx 1.1 \times 10^{17}$ cm. Since $R_H \propto t_d$, this is the smallest distance in this scenario. About half of the flux changes between consecutive 1-day bins in F1 have doubling or halving times of 0.5--2 days, which correspond to $R_H \approx (1.7$--$6.7) \times 10^{17}$ cm for $\Gamma = 12$. These estimates are compared with the SED results in Section~\ref{subsec:sed}.

\begin{deluxetable}{ll}[ht]
%\digitalasset
\tablewidth{0pt}
\tablecaption{Fractional variability in different wavebands from MJD 59215 to MJD 60751. \label{tab:f_var}}
\tablehead{
\colhead{\textbf{Waveband}} & \colhead{\textbf{F$_{var}$}}
}
\startdata
$\gamma$-ray & 1.20 $\pm$ 0.01 \\ \hline
X-ray & 0.20 $\pm$ 0.06 \\ \hline
UVOT-V & 0.44 $\pm$ 0.01 \\ \hline
UVOT-B & 0.45 $\pm$ 0.01 \\ \hline
UVOT-U & 0.42 $\pm$ 0.01 \\ \hline
UVOT-W1 & 0.40 $\pm$ 0.01 \\ \hline
UVOT-W2 & 0.56 $\pm$ 0.01 \\ \hline
UVOT-M2 & 0.43 $\pm$ 0.01 \\ \hline
ASAS-SN g & 0.63 $\pm$ 0.01 \\ \hline
ZTF g & 1.16 $\pm$ 0.01 \\ \hline
ZTF r & 1.30 $\pm$ 0.01 \\ \hline
NeoWISE W1 & 1.16 $\pm$ 0.01 \\ \hline
NeoWISE W2 & 1.13 $\pm$ 0.01 \\ \hline
\enddata
\end{deluxetable}

\subsection{Correlation study}
\begin{figure}
    \centering
    \includegraphics[width=1\linewidth]{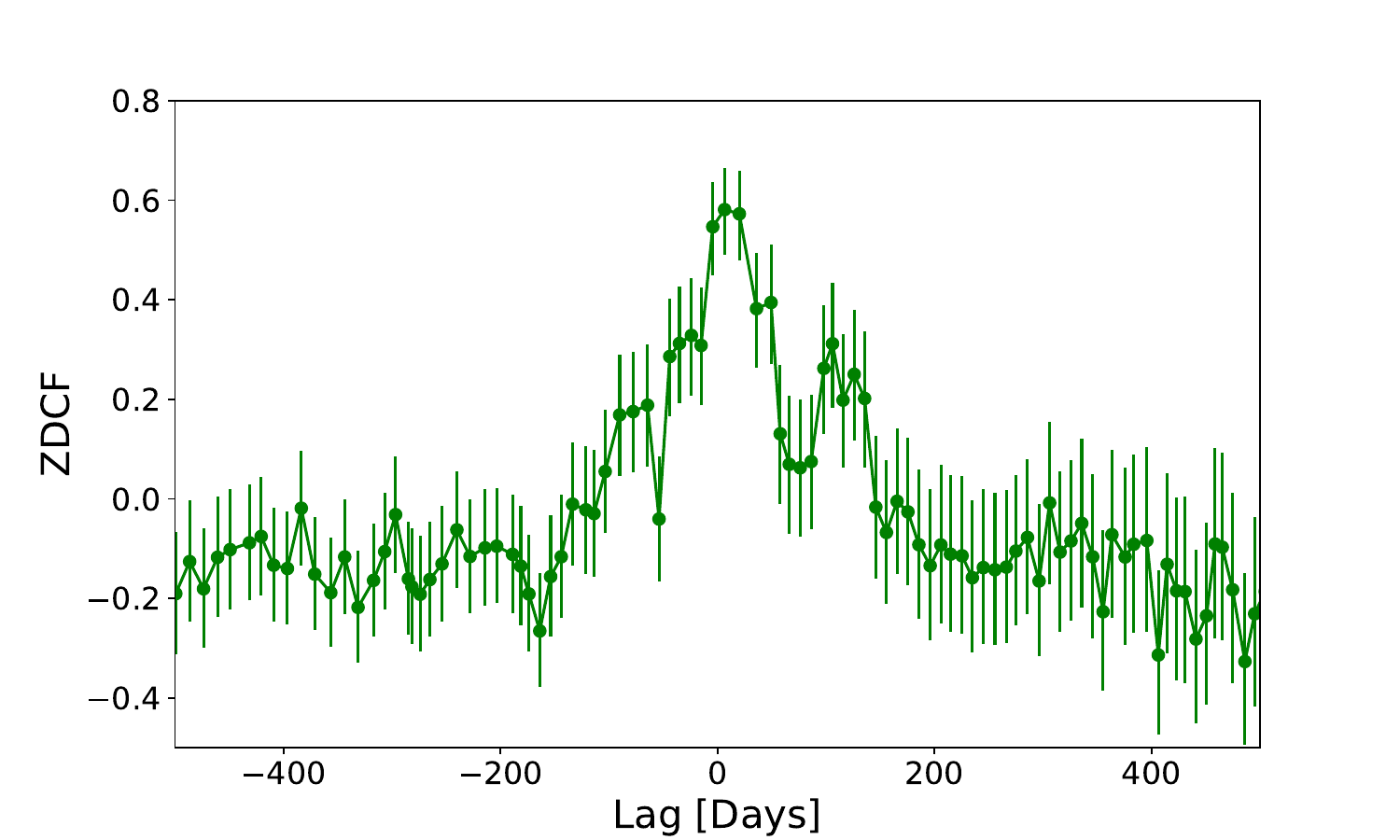}
    \caption{The cross-correlation function estimated between $\gamma$-ray and optical ASASSN g-band using zDCF.}
    \label{fig:zdcf_gamma_opt}
\end{figure}

To find correlation in different wave bands, we have adopted the z-transformed discrete correlation function (ZDCF; \citealt{1997ASSL..218..163A,2014ascl.soft04002A}). We have performed zDCF analysis between various wavebands. The good sampling in the Fermi-LAT $\gamma$-ray and ASAS-SN g-band light curves made it possible to derive the correlation among them. We found a peak with a time lag of $6.52^{+7.95}_{-1.32}$ days and a correlation coefficient above 0.6 (see Fig~\ref{fig:zdcf_gamma_opt}), suggesting a good correlation between $\gamma$-ray and optical wave bands. As the binning time of the $\gamma$-ray light curve is 10 days, at any time lag calculated by zDCF below 10 days can be considered as 'no time lag'. As we have fewer observations in the X-ray, UV, and IR bands, we did not find any significant correlation between them using zDCF. But by visual inspection from Figure \ref{fig:all_lcs}, we can see simultaneous flaring in all the wave bands. These results suggest that the emissions are produced simultaneously at the same location, and the same population of electrons is responsible for it.
\citet{10.1093/mnras/stad060} have studied an optical-$\gamma$-ray correlation using ASAS-SN and Fermi-LAT data for a large number of objects, and they found that a large number of blazar shows strong optical-$\gamma$-ray correlation with an average time lag of 1.1$^{+7.1}_{-8.5}$ day irrespective of FSRQs, BL Lacs objects or low, intermediate, and high synchrotron peak blazar. \citet{2016Galax...4...64W} also have shown that the average time lag between $\gamma$-ray and optical for different blazars is 0 $\pm$ 7 days.
Our result is consistent with their result, and thus, we consider the co-spatial origin of emissions in different wave bands,  which leads to the consideration of a one-zone leptonic emission model for broadband SED modeling.

\subsection{Spectral variations}
\label{subsec:spectral_var}
\begin{figure}
    \centering
    \includegraphics[width=0.99\linewidth]{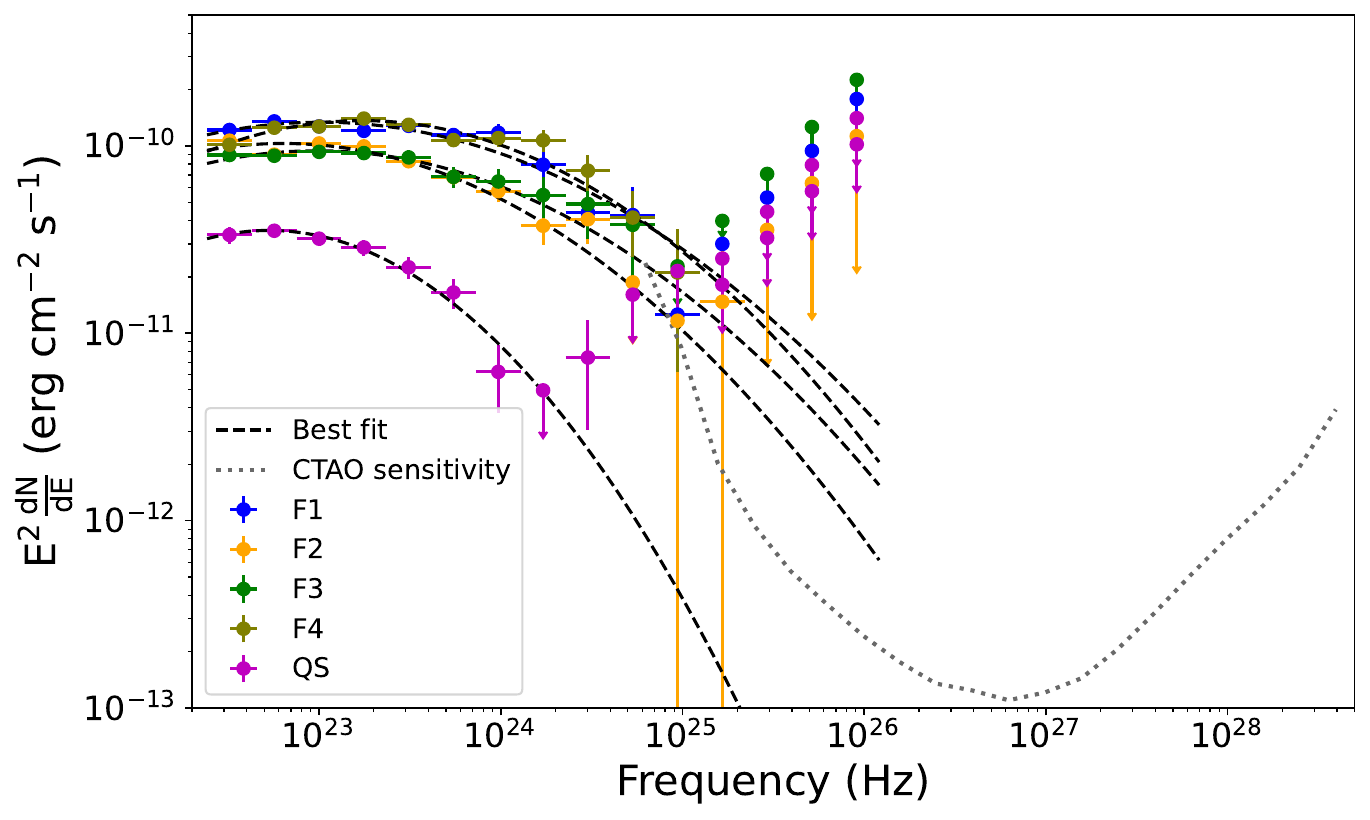}
    \caption{The $\gamma$-ray spectrum for F1, F2, F3, F4, and (quiet state) QS fitted with log-parabola model. The data point with the arrow represents the upper limit. The frequency is in the observed frame. The gray dotted line represents the CTAO-North (50h) differential sensitivity.}
    \label{fig:gamma_seds}
\end{figure}

The $\gamma$-ray spectra for each of the states of 4C +27.50 have been plotted in Figure \ref{fig:gamma_seds}, and the frequency is in the observed frame. The spectra were fitted with a log-parabola (LP) model. The LP model is defined as
\begin{equation}
    \frac{dN(E)}{dE} = K\left(\frac{E}{E_{\text{break}}}\right)^{-\alpha-\beta\,\text{log}\left(\frac{E}{E_{\text{break}}}\right)}
    \label{eq:lp}
\end{equation}
Here, $\frac{dN(E)}{dE}$ is the number of photons in erg$^{-1}$ cm$^{-2}$ s$^{-1}$, $K$ is the normalization, $\alpha$ is the photon index at $E_{\text{break}}$ and $\beta$ is the curvature index that determines the curvature around the $E_{\text{break}}$.

\citet{2022MNRAS.515.4505M} showed that 4C +27.50 can be marginally detected by CTAO (Omega and Alpha configurations). We plotted the CTAO-North 30-degree (50h) differential sensitivity with a grey dotted line in Figure \ref{fig:gamma_seds}. The sensitivity curve of CTAO overlaps with Fermi-detected upper limits. Thus, there is a possibility of detection by CTAO in the near future.

The $\gamma$-ray SEDs of F1, F2, and F3 are situated well above the SED of QS, justifying higher energetic $\gamma$-ray emission during flaring periods. The log-parabola fitted parameters of the spectra are tabulated in Table \ref{tab:gamma_sed}. The best-fit spectral index derived for the quiet state is much steeper than the other flaring states. The curvature parameter for the quiet state is also higher, suggesting a significantly curved spectrum. It is evident in all the cases that the spectrum finishes before 100 GeV, suggesting that no high-energy photon above 100 GeV is observed in the observer frame. 
%This information is crucial to constrain the location of the emission site. As suggested in various studies, the emission sites located within the broad-line region can not produce gamma-ray photons above 10 GeV due to the high pair production gamma-ray opacity \cite{2006ApJ...653.1089L}. Therefore, we consider a one-zone leptonic emission model to explain the broadband SEDs by locating the emission site within the BLR. 

\begin{figure*}[ht]
    \centering
    \subfloat[]{\includegraphics[width=0.48\textwidth]{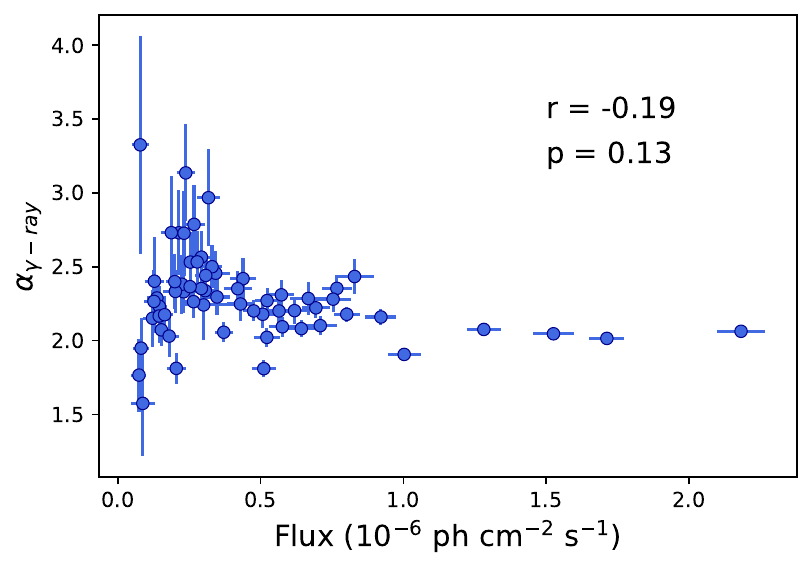}}
    \hfill
    \subfloat[]{\includegraphics[width=0.48\textwidth]{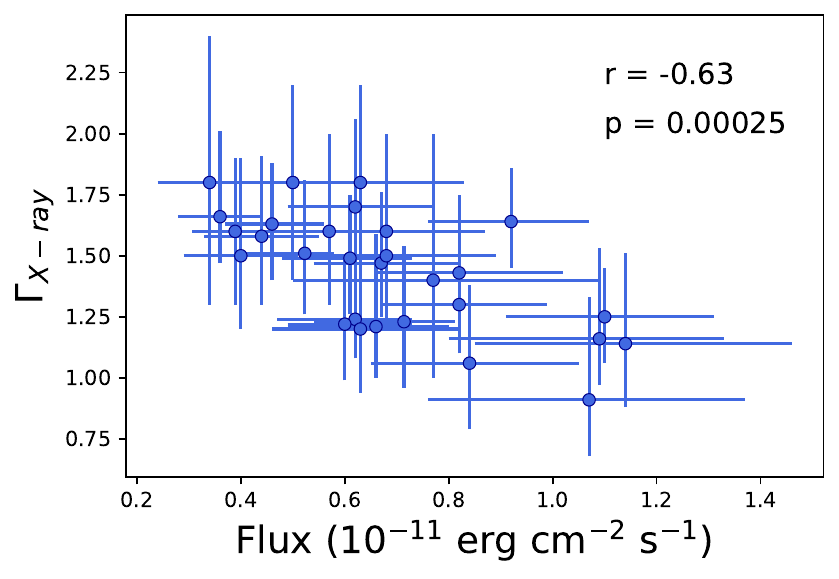}}
    \hfill
    \subfloat[]{\includegraphics[width=0.48\textwidth]{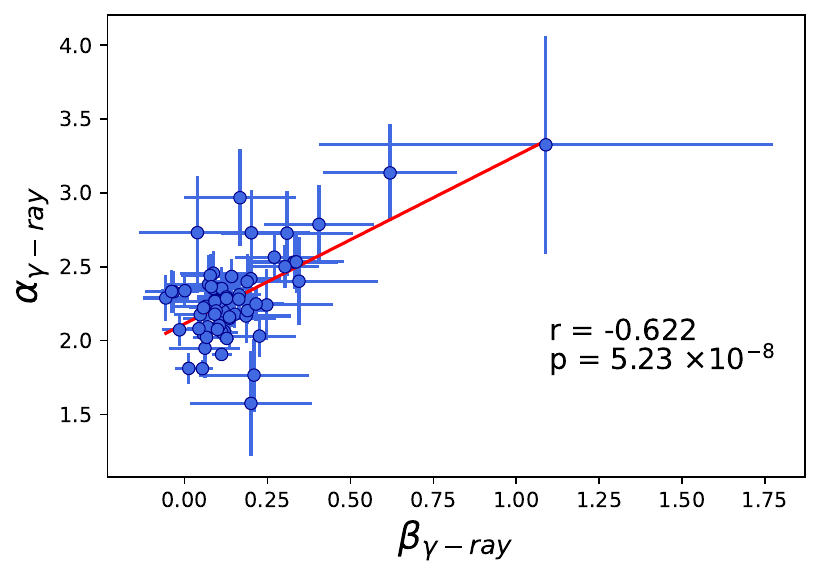}}
    \caption{(a) Variation of $\gamma$-ray flux with $\gamma$-ray index. (b) Variation of X-ray flux with X-ray index. (c) Variation of $\alpha_{\gamma-ray}$ with the curvature in $\gamma$-ray spectra ($\beta_{\gamma-ray}$). The red line shown in (c) indicates the best-fit straight line.}
    \label{fig:spectral_var}
\end{figure*}

The curvature in the photon spectrum can also be produced by various other reasons. The blazar emission can be considered as a two-step process, namely acceleration and radiation via various cooling processes. Earlier studies suggest that as soon as the particle is injected into the jet, they get accelerated under a shock and produce a power-law particle distribution (Fermi-1st order) spectrum, and once they achieve a certain amount of energy, they like to leave the shock and go through various cooling processes such as synchrotron and inverse-Compton, and produces the curvature in the photon spectra. On the other hand, \cite{2004A&A...413..489M, 2006A&A...448..861M} argued that the spectral curvature observed in the photon spectrum could be the intrinsic feature of the particle spectrum, and the log-parabola particle distribution can be obtained by a simple model for statistical acceleration. As suggested in \cite{2004A&A...413..489M}, we also plot the log-parabola spectral index and the curvature parameter in Fig~\ref{fig:spectral_var}(c). A clear positive correlation is seen with Pearson's correlation coefficient $r = 0.62$ and P-value = 5.24$\times$10$^{-8}$, suggesting the hypothesis that the curvature is caused by the particle acceleration process. \citet{2018MNRAS.477.4749C} found evidence that in most of the blazars, $\gamma$-rays are produced outside the BLR region and the $\gamma$-ray spectrum of Fermi detected blazars are mostly intrinsic. Similarly, \citet{2019ApJ...877...39M} corroborates the production of $\gamma$-rays outside the BLR region with a smaller sample of blazars.

\begin{deluxetable}{lll}[ht]
%\digitalasset
\tablewidth{0pt}
\tablecaption{Log-parabola fitted parameters of $\gamma$-ray SEDs for the different states. \label{tab:gamma_sed}}
\tablehead{
\colhead{\textbf{State}} & \colhead{\textbf{$\alpha$}} & \colhead{\textbf{$\beta$}}
}
\startdata
F1 & 2.11 $\pm$ 0.02 & 0.07 $\pm$ 0.01 \\ \hline
F2 &  2.22 $\pm$ 0.02 & 0.09 $\pm$ 0.02 \\ \hline
F3 &  2.13 $\pm$ 0.03 & 0.08 $\pm$ 0.02 \\ \hline
F4 &  2.04 $\pm$ 0.02 & 0.1 $\pm$ 0.01 \\ \hline
QS & 2.46 $\pm$ 0.06 & 0.16 $\pm$ 0.04 \\ \hline
\enddata
\end{deluxetable}

We also investigated the flux index correlation in gamma-rays and X-rays. We used the X-ray and gamma-ray light curves from Figure \ref{fig:all_lcs} and their corresponding spectral index. The 10-day binned gamma-ray light curve was produced for the log-parabola spectral model, and therefore, we have the photon spectral index ($\alpha_{\gamma-ray}$). Whereas, the X-ray light curves were derived by fitting the power-law spectral model, and therefore, we have $\Gamma_{X-ray}$. We performed a Spearman's rank correlation analysis to examine the possible correlation between flux and photon spectral index in both cases.
Scatter plot between $\gamma$-ray spectral index ($\alpha_{\gamma-ray}$) and $\gamma$-ray flux has been shown in Figure \ref{fig:spectral_var} (a). The Spearman’s rank correlation analysis resulted in a rank coefficient of $r=-0.19$, with a null-hypothesis probability of $P=0.13$. Thus high value of $P$ indicates the correlation between $\gamma$-ray flux and $\alpha_{\gamma-ray}$ is indecisive.
On the other hand, Spearman’s rank correlation analysis between X-ray flux and $\alpha_{X-ray}$ results in a correlation coefficient $r=-0.63$ and $P=0.00025$, suggesting a high anti-correlation as shown in Figure \ref{fig:spectral_var} (b). It suggests a harder-when-brighter trend. A similar trend of $\gamma$-ray and X-ray is also seen in other blazars such as Ton 599 \citep{2024arXiv241023194M}.
%A linear relation has been observed between $\alpha_{\gamma-ray}$ and curvature in $\gamma$-ray spectra ($\beta_{\gamma-ray}$) as shown in Figure \ref{fig:spectral_var} (c).

\subsection{Broadband SED modeling}
\label{subsec:sed}
This is the first time we have compiled the broadband information about this source and performed the SED modeling. The averaged UVOT image was produced using the task \texttt{uvotimsum}, and the flux used for broadband SED was calculated using the procedure stated in Section \ref{subsec:UVOT}. The averaged flux value of the archival data from ZTF and ASAS-SN was used in the broadband SED. The combined Swift XRT data were used in broadband SED following the procedure mentioned in Section \ref{subsec:XRT}. The Fermi LAT spectra shown in Figure \ref{fig:gamma_seds} were used in the individual broadband SED. Archival SED data were compiled from the \texttt{SSDC SED Builder}\footnote{\url{https://tools.ssdc.asi.it/SED/}} tool.
Broadband SED modeling helps to probe the emission mechanism happening in jets of blazars and also reveals the physical scenario happening close to the SMBH. We modeled the broadband SED of QS, F1, F2, and F3 states using the publicly available code \texttt{JetSeT}\footnote{\url{https://jetset.readthedocs.io/}} \citep{2020ascl.soft09001T,2011ApJ...739...66T,2009A&A...501..879T}. \texttt{JetSeT} fits numerical models to the data points to provide the optimal jet parameters.

In general, the SED modeling involves a large number of parameters, and sometimes the model is degenerate. To avoid any kind of degeneracy and to reduce the number of free parameters in the model, it is important to fix some of the parameters from the observations for better fit results. From section \ref{subsec:var_study}, the observed variability time constrained the size of the emission region to $R \lesssim 8.9 \times 10^{15}$ cm and, in the colliding-shell scenario, its location down the jet to $R_H \gtrsim 1.1 \times 10^{17}$ cm. For the source $\delta = 23.9^{+21.09}_{-12.12}$ and viewing angle ($\theta$) is in the range $0.01<\theta<2.48$ derived in \cite{2018ApJ...866..137L}. For the first flaring state, we made free the $\theta$ between 0.01$^\circ$ and 2.48$^\circ$, and it came to be 0.5$^\circ$ for the best fitting. Therefore, for SED modeling, we fixed $\theta$ to 0.5$^\circ$ for the rest of the epochs. The luminosity of the disk ($L_{disk}$) is fixed to the value of $10^{46}$ erg s$^{-1}$ taken from \cite{2021ApJS..253...46P} for the flaring states. The QS fit was performed with $L_{disk} = 10^{45}$ erg s$^{-1}$ (Table~\ref{tab:sed_params}). We kept $R_{BLR-in}$, $R_{BLR-out}$, and $R_{DT}$ at the values for $L_{disk} = 10^{46}$ erg s$^{-1}$ in all states. The sizes of the BLR and the DT are set by the long-term disk luminosity and can follow a change of luminosity only on their light-crossing times, $R_{BLR-in}/c \approx 4$ months and $R_{DT}/c \approx 8$ yr in the source frame, much longer than the 30 days (observed frame) that separate the QS from F1. In the QS fit, $L_{disk}$ therefore sets only the energy densities of the external photon fields. The disk luminosity was not measured during the QS, so the lower $L_{disk}$ should be read as a weaker external photon field in the QS, not as a measured dimming of the disk. As 4C +27.50 is classified as an FSRQ, we consider the existence of the BLR. The BLR is composed of various gaseous clouds in a Keplerian orbit around the central black hole. These clouds get irradiated by the central source and produce the broad emission lines. The Ly$\alpha$ emission line (at approximately 1216~\AA\ in the rest frame) is the strongest and most important spectral feature in the ultraviolet spectrum of the BLR \citep{2001AJ....122..549V,2006agna.book.....O}. We therefore approximate the BLR radiation field by the Ly$\alpha$ line alone, treated as a monochromatic (delta-function) photon field in \texttt{JetSeT}, and neglect the other broad lines.

Due to the distribution of the clouds, the BLR is assumed to have a shell-type structure with other inner and outer radii linked with the accretion disk luminosity by \citep{2009MNRAS.397..985G}
\begin{equation}
    R_{BLR-in} = 10^{17}\times\sqrt{\frac{L_{disk}}{10^{45}~erg/s}} ~cm
\end{equation}
and the outer radius of BLR, $R_{BLR-out}$ = $1.1 \times R_{BLR-in}$ \citep{2024A&A...689A..56E}. Thus we found $R_{BLR-in}\approx3.16\times10^{17}$ cm and $R_{BLR-out}\approx3.48\times10^{17}$ cm. The colliding-shell estimate for the shortest variability ($R_H \approx 1.1\times10^{17}$ cm) lies inside the BLR, while variability on 0.5--2 day time-scales places the emission region at $(1.7$-$6.7)\times10^{17}$ cm, from inside the BLR to about twice $R_{BLR-out}$. We calculated the radius of the dusty torus by \citep{2011MNRAS.414.2674G}
\begin{equation}
    R_{DT} = 2.5 \times 10^{18}\times\sqrt{\frac{L_{disk}}{10^{45}erg/s}} ~ cm
\end{equation}
and found $R_{DT} = 7.91\times 10^{18}$ cm. It is also important to note that at this distance, the BLR photons or DT photons do not contribute much to the absorption of gamma-ray photons due to the pair-production process. This is much relevant when the emission region is well buried in the BLR and is located close to the base of the jet.
We can estimate the peak frequency corresponding to the black hole mass and disk luminosity using \citep{2013MNRAS.431..210C}
\begin{equation}
    \frac{\text{M}}{10^9 \text{M}_{\odot}} = 1.44 \left(\frac{\nu_p}{10^{15} \text{Hz}}\right)^{-2}\left(\frac{L_{disk}}{<2\text{cos}\theta>\times 10^{45} \text{erg/s}}\right)^{1/2}.
\end{equation}
Taking $M=10^{8.8}M_{\odot}$, we found peak frequency $\nu_p = 2.33\times10^{15}$ Hz. Assuming perfect blackbody emission, Wien's displacement law gives the disk temperature $T_{disk}$ = 3.97$\times10^4$ K. Thus, during the broadband SED fitting, we fixed the values of $z$, $R_{BLR-in}$, $R_{BLR-out}$, $R_{DT}$, $\theta$, $T_{disk}$, and $L_{disk}$ to reduce the number of free parameters. We keep $R$ and $R_H$ as free parameters and later compare them with the estimated values. We have chosen the temperature of the dusty torus ($T_{DT}$) to be 1210K, which is a standard value for $\gamma$-ray bright blazars \citep{2011ApJ...732..116M}.

\begin{figure*}[ht]
    \centering
    \subfloat[]{\includegraphics[width=0.49\textwidth]{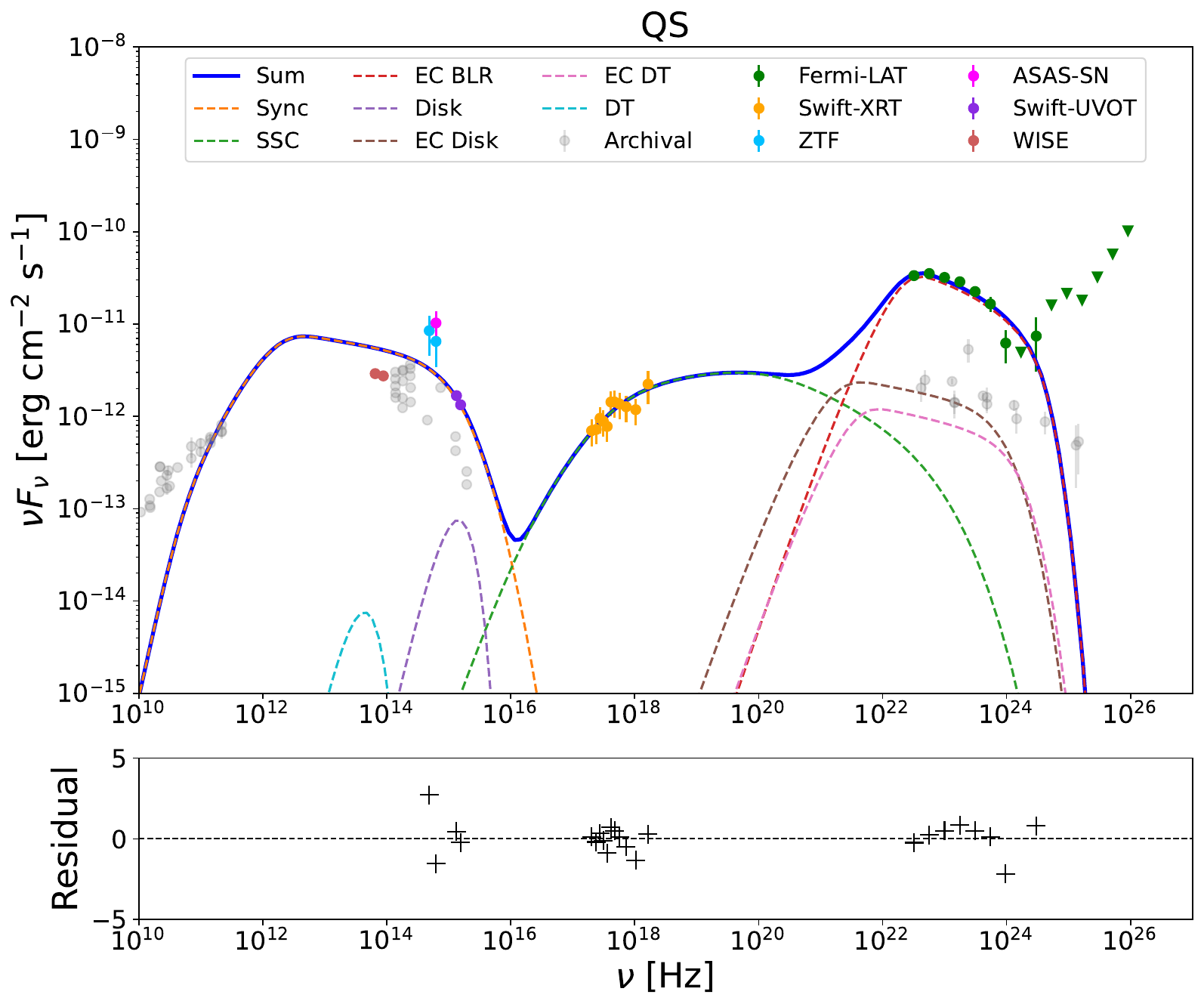}}
    \hfill
    \subfloat[]{\includegraphics[width=0.49\textwidth]{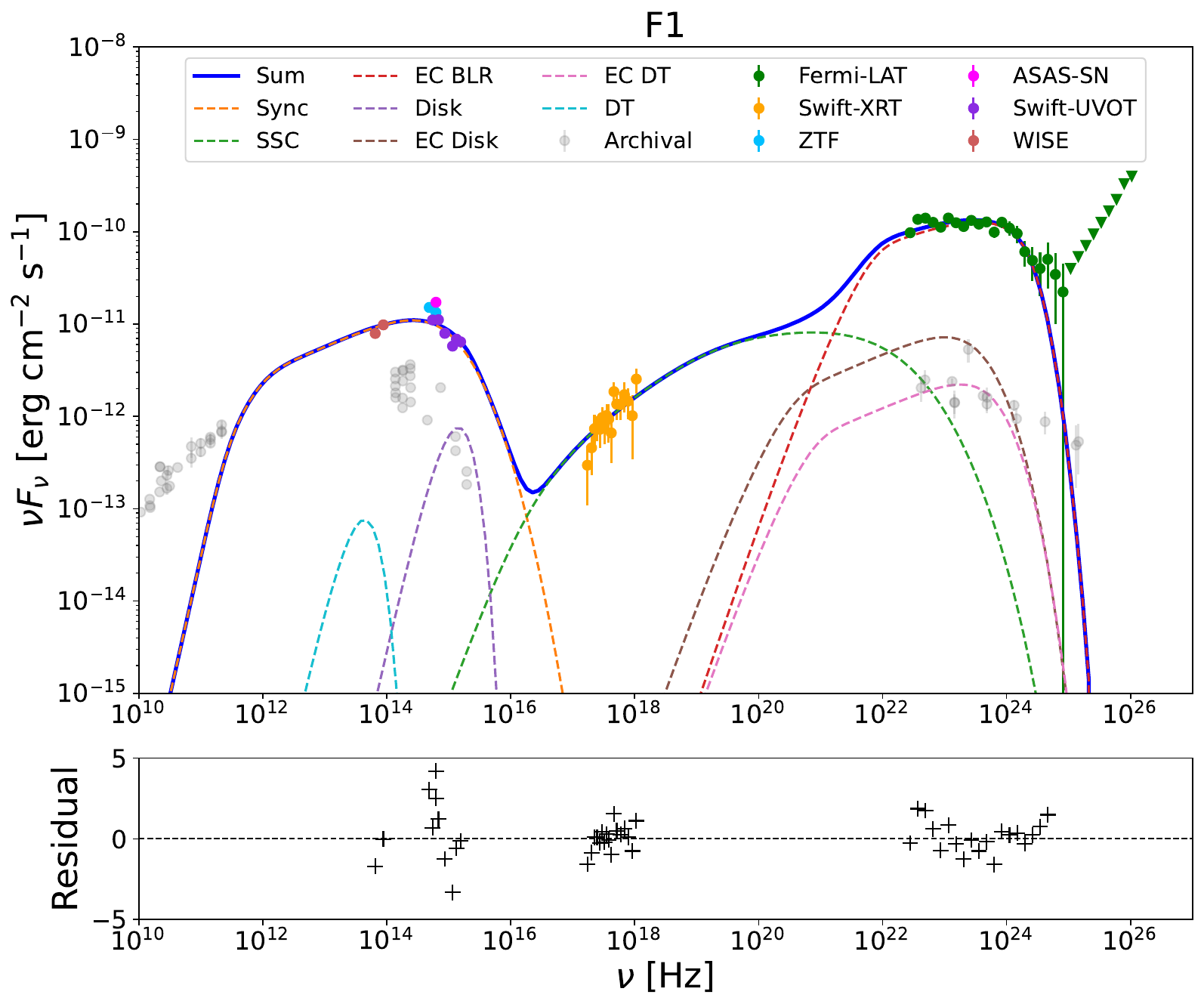}}

    \subfloat[]{\includegraphics[width=0.49\textwidth]{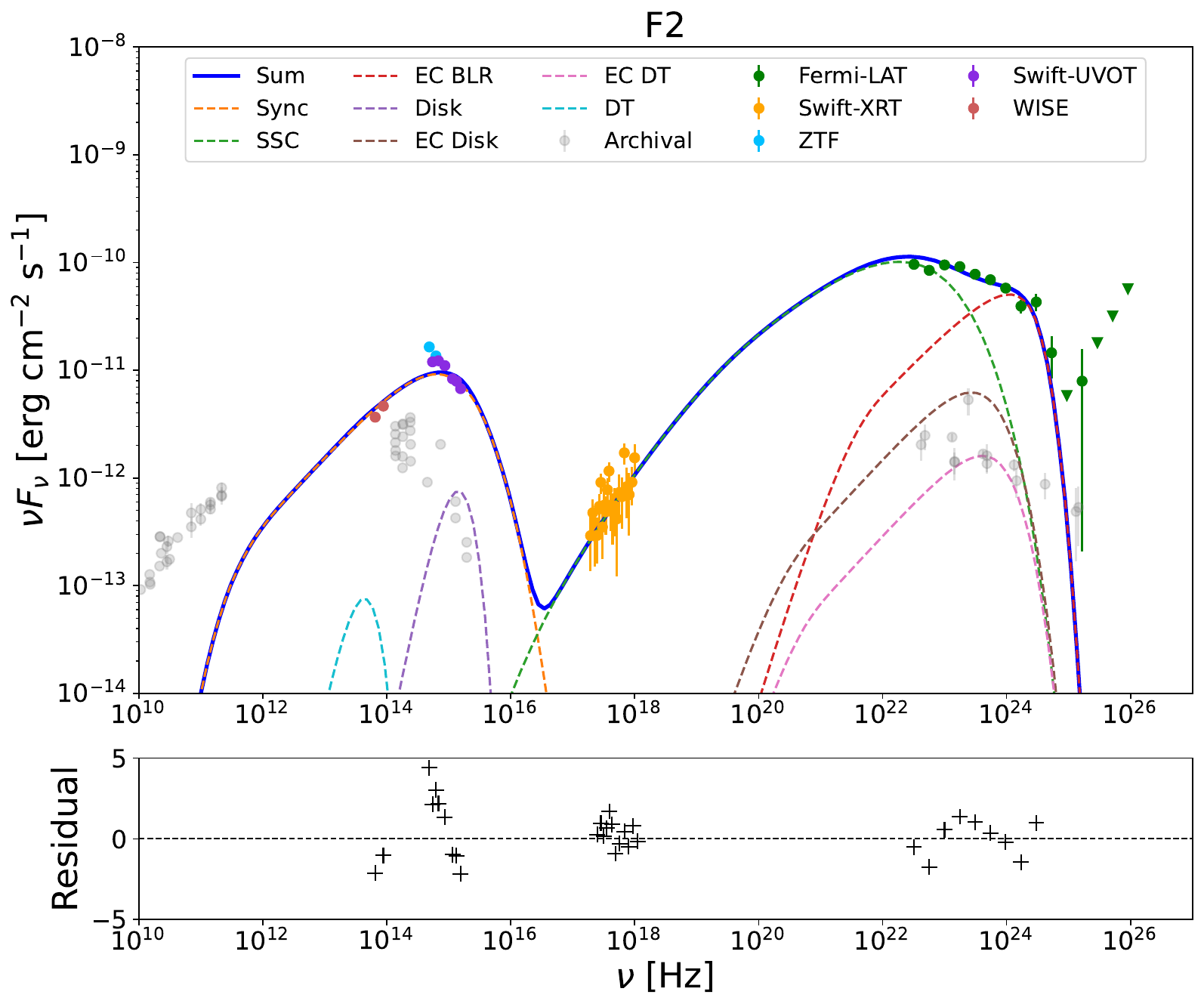}}
    \hfill
    \subfloat[]{\includegraphics[width=0.49\textwidth]{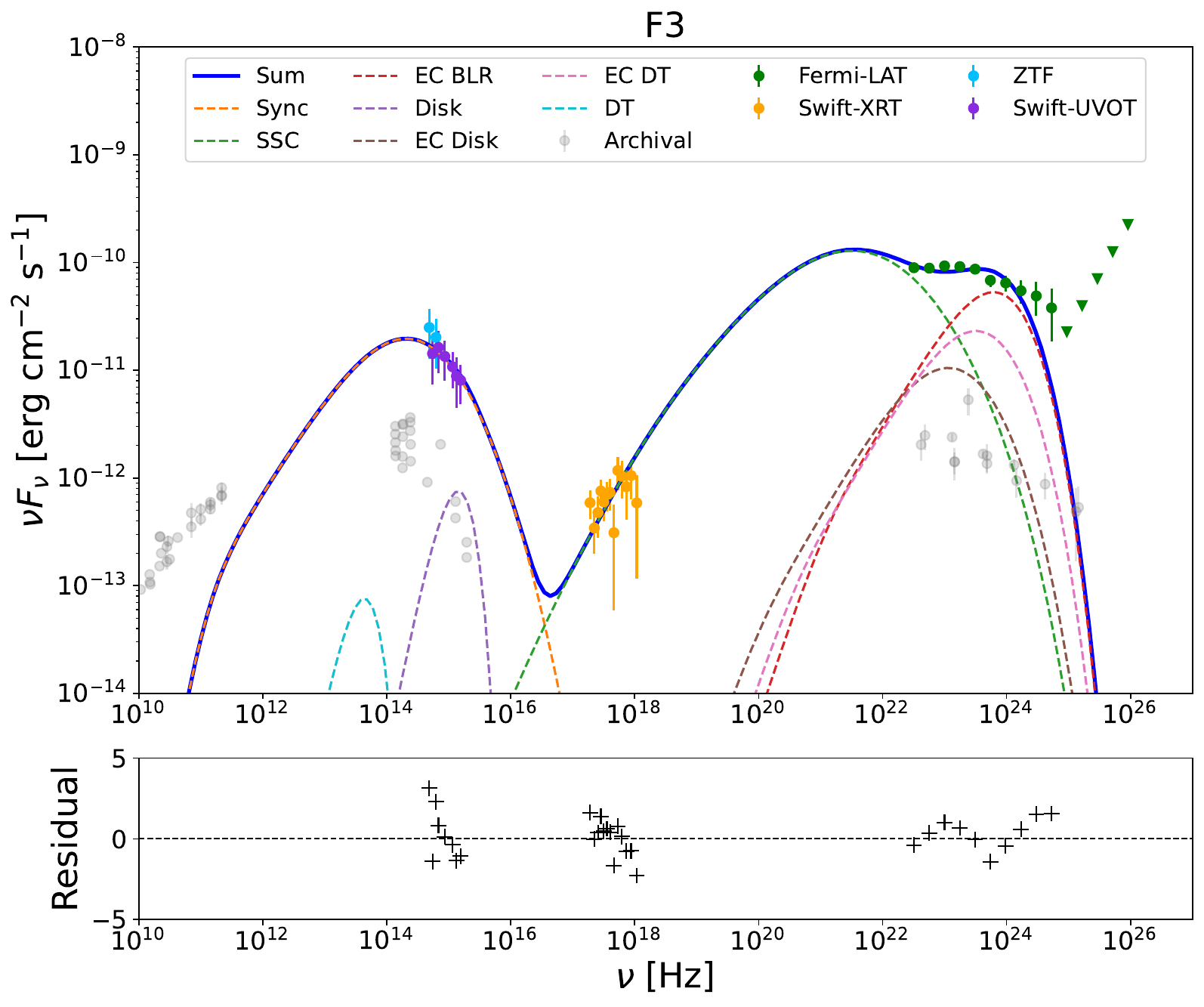}}
    \caption{The broadband SEDs of the QS (a), F1 (b), F2 (c), and F3 (d) fitted with one-zone leptonic model. The data and the various colorful lines are self-explanatory. The grey data points shown in all the upper panels are archival data to guide the fit. The downward green triangles in the upper panels represent the upper limits of detection in $\gamma$-rays. The lower panel shows the residuals of the best-fit model. The error bars of the UVOT, ZTF, ASAS-SN, and WISE points include the standard deviation of the fluxes within each window, added in quadrature to the measurement error.}
    \label{fig:sedfig}
\end{figure*}

Following the shape of the $\gamma$-ray SED, we adopted a log-parabola electron distribution with a low-energy power-law branch (the \texttt{lppl} distribution of \texttt{JetSeT}),
\begin{equation}
n(\gamma) = N K \left\{
\begin{array}{ll}
(\gamma/\gamma_0)^{-\alpha}, & \gamma_{min} \leq \gamma \leq \gamma_0 \\
(\gamma/\gamma_0)^{-\alpha-\beta\,\log(\gamma/\gamma_0)}, & \gamma_0 < \gamma \leq \gamma_{max}
\end{array}
\right.
\label{eq:lppl}
\end{equation}
where $N$ is the electron number density, $K$ normalizes the integral of $n(\gamma)/N$ to unity, $\alpha$ is the low-energy spectral index, $\beta$ is the curvature and $\gamma_0$ is the turn-over Lorentz factor. The photon flux resulting from synchrotron and various cases of inverse-Compton scattering is determined by the \texttt{JetSeT} using this particle distribution by solving the transport equations \citep{2011ApJ...739...66T}. The free parameters of the model are particle distribution, magnetic field, Lorentz factor, $R$, and $R_H$, which are optimized using \texttt{minuit} optimizer \citep{2010JPhCS.219d2044L} after adding a 10\% systematic uncertainty to all data points, and best-fit values are tabulated in Table \ref{tab:sed_params} and the best-fit SEDs are shown in Figure \ref{fig:sedfig}. We have applied extragalactic background light (EBL) correction to the broadband SED using the \texttt{EBLAbsorptionTemplate} \citep{2008A&A...487..837F} in \texttt{JetSeT}. We have fixed the parameters related to the external photon fields, such as BLR, Disk, and DT, as their structure do not change significantly for a short period. All three fields are computed by \texttt{JetSeT}. The disk is a single-temperature blackbody at $T_{disk}$ with luminosity $L_{disk}$. The BLR is a spherical shell between $R_{BLR-in}$ and $R_{BLR-out}$ that reprocesses a fraction $\tau_{BLR}$ of $L_{disk}$ into the Ly$\alpha$ line, following \citet{2003APh....18..377D}. The DT is a uniform blackbody field at $T_{DT}$ inside $R_{DT}$, with luminosity $\tau_{DT}\,L_{disk}$. The energy densities $U_{Disk}$, $U_{BLR}$, and $U_{DT}$ in Table~\ref{tab:sed_params} are evaluated at $R_H$ and given in the comoving (blob rest) frame.

SED modeling suggested that in all four epochs, the emission region lies just outside the BLR ($R_H = 3.6$-$5.4\times10^{17}$ cm, compared with $R_{BLR-out} = 3.48\times10^{17}$ cm) and well inside the DT. The emission region size is found to be much larger in QS compared to other flaring states.

%Along with this, we also have parameters related to external photon fields such as BLR, disk, and DT temperature, and their energy density is kept free to quantify their contribution. 
Since the emission region is located closer to BLR, as expected, BLR photons contribute more to inverse-Compton, and hence the BLR energy density is higher than the disk and the dusty torus. Across all the states, the disk and dusty torus contribution is very less compared to BLR photons. We also observed a change in BLR photon density during different flux states. During the quiet state, it is estimated as 0.03 erg/cm$^3$, and it increases to 1.5 and 0.50 erg cm$^{-3}$ in F1 and F2, and to 0.045 erg cm$^{-3}$ in F3, where $R_H$ is largest. Part of the low QS value comes from the ten times lower $L_{disk}$ used in the QS fit.

The low-energy spectral index of the particle distribution is found to be softer in the quiet state compared to all flaring states. This implies the presence of more high-energy electrons in the jet. The harder spectral index during flaring states presence of an efficient particle acceleration mechanism for flares to take place. The magnetic field is derived as 0.39 G in the quiet state and has increased to nearly 1 G in flaring states, which is consistent with the typical value of the magnetic field in blazars \citep{10.1111/j.1365-2966.2009.15397.x}. The $\gamma_{max}$ in this source is found to be higher (10$^{5}$) compared to the typical value for FSRQs, which is around 10$^{3-4}$ \citep{10.1111/j.1365-2966.2009.15007.x, 10.1111/j.1365-2966.2009.15397.x}.

The bulk Lorentz factor, $\Gamma$, estimated in these flaring states is around 10, which is similar to the typical FSRQs as seen in \cite{10.1111/j.1365-2966.2009.15397.x}. We found a size of the emission region as  6.2$\times$10$^{16}$ cm for the quiet state, which is larger compared to the flaring state's size of the emission region (between 0.79--1.4$\times$10$^{16}$ cm), as expected since the variability is slower during the quiet state. With the Doppler factor of each state (Table~\ref{tab:sed_params}), the 7.8 h doubling time gives $R \leq (6.3$--$9.5)\times10^{15}$ cm for the flares, and the fitted sizes exceed these limits by factors of 1.2--1.6. The fitted sizes correspond to variability times of 9--13 h, comparable to the 1-day bins from which the 7.8 h value was derived. The 7.8 h doubling occurred once, between two consecutive bins, whereas each flare SED averages 65--85 days of emission. Likewise, Equation~\ref{eq:rh} with the fitted $\Gamma$ (8.5--13) gives $R_H = (0.5$--$1.3)\times10^{17}$ cm for $t_d = 7.8$ h, three to seven times smaller than the fitted $R_H$. The fitted $R_H$ values correspond to $t_d \approx 1$--2 days, the time scale of about half of the flux changes in the 1-day binned light curve of F1. This does not mean that Equation~\ref{eq:rh} or the fits are wrong. Equation~\ref{eq:rh} links a variability time to a dissipation distance, so in the colliding-shell picture faster variations arise closer to the black hole and slower ones farther out. Each SED averages 65--85 days and is dominated by the day-scale variations, for which Equation~\ref{eq:rh} gives the fitted $R_H$. A one-zone model cannot describe the time-averaged SED and the single fastest event at the same time, and the 7.8 hrs event most likely came from a more compact region that contributes little to the time-averaged SED.

We define the jet powers as $L_i = \pi R^2 \Gamma^2 c\, U_i$, where $U_i$ is the comoving energy density of relativistic electrons ($L_e$), cold protons ($L_p$, one proton per ten electrons), or magnetic field ($L_B$). The kinetic power is $L_{kin} = L_e + L_p$, the radiative power is $L_{rad} = \Gamma^2 L'_{rad}/4$, with $L'_{rad}$ the comoving radiative luminosity, and the total jet power is $L_{Jet} = L_{rad} + L_{kin} + L_B$. The ratio $L_e/L_B$ is 2.1, 2.0, 31, and 40 for QS, F1, F2, and F3, so the jet is particle-dominated in all states, close to equipartition in QS and F1. The model reproduces the LAT spectra of all states within about $2\sigma$, with an rms of the $\gamma$-ray residuals of about $1\sigma$, and the X-ray spectra with an rms of 0.6--1.1$\sigma$. Above 10 GeV (observed frame) the model lies slightly below the data, by 0.2--1.6$\sigma$. This small deficit could come from a harder high-energy tail of the electron distribution than the assumed log-parabola, or from an additional hadronic component, but the present data do not require either.

The small deficit of the model at the highest LAT energies (Figure \ref{fig:sedfig}) could potentially be reduced by considering the reduced efficiency of electron cooling in the Klein-Nishina regime. As demonstrated by \citet{2005MNRAS.363..954M}, in radiation-dominated environments, the suppression of the Compton scattering cross-section leads to a hardening or upturn in the high-energy tail of the steady state electron distribution. This population of excess high-energy electrons could sustain the $\gamma$-ray flux at energies where standard Thomson-cooling models predict a steeper cutoff. For instance, \citet{2018ApJ...853....6L} demonstrates that solving the particle transport equation with stochastic acceleration and full Klein-Nishina cross-sections yields electron distributions that are harder than standard ad-hoc functions and successfully reproduce high-energy $\gamma$-ray data. On the other hand, a hadronic component can also possibly explain the high-energy part with any further detection of very high energetic (VHE) $\gamma$-rays. As discussed in Section \ref{subsec:spectral_var}, detection of VHE using CTAO is possible, and CTAO detection will help to constrain the VHE part of the broadband SED.

The WISE magnitudes were converted to $\nu F_\nu$ using the effective frequencies of the W1 and W2 bands, $8.82\times10^{13}$ and $6.52\times10^{13}$ Hz. The model also misses some IR, optical, and UV points. These data are not simultaneous within each window, and the source varies strongly in these bands (Figure~\ref{fig:all_lcs}). The fractional standard deviation of the fluxes within the windows is 28--49\% for UVOT in F2 and F3, 36--72\% for ZTF, and 35--50\% for ASAS-SN. The two UVOT visits in F1 differ by 0.5--1.9 mag, and QS has a single UVOT visit. NeoWISE observed the source in a single visit of a few days during each of QS, F1, and F2, and not during F3, after the end of the NeoWISE mission. In the QS, which was fitted without the WISE points, the model lies a factor of about two above this single visit. The variability within the windows is larger than the 10\% systematic uncertainty used in the fit, and the measured scatter is only a lower limit because variability between the visits was not sampled. Part of the mismatch thus reflects variability within each window rather than a single steady state. The largest residuals of the fits are in the optical-UV, up to $4.4\sigma$. When the scatter within each window is added to the uncertainties, the rms of the optical-UV residuals falls to 0.3--1.2$\sigma$, so the one-zone model describes these data within their variability.

\begin{figure}
    \centering
    \includegraphics[width=0.99\linewidth]{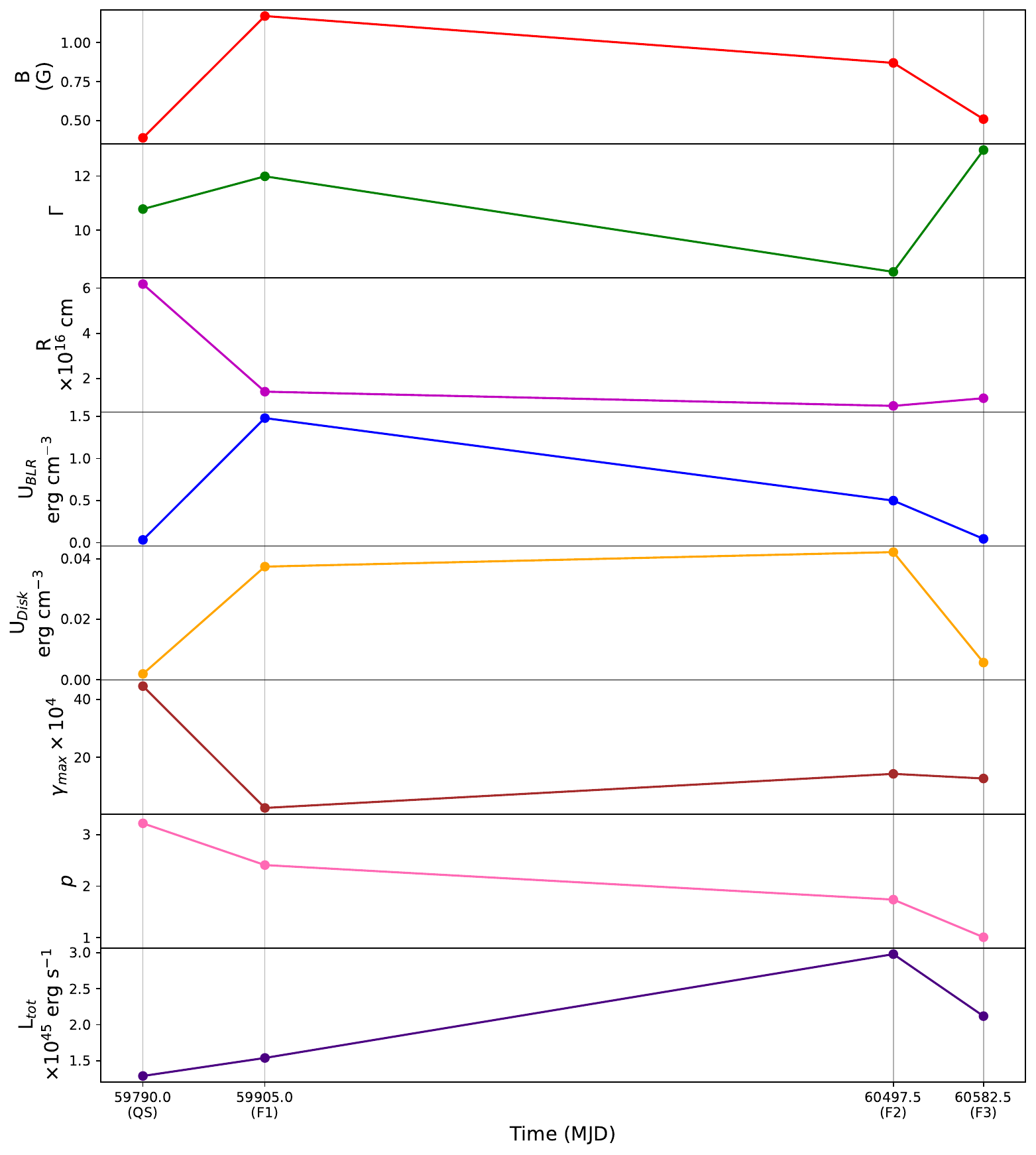}
    \caption{Variation of jet parameters in different epochs.}
    \label{fig:jet_param_ephocs}
\end{figure}

We have also summarized the variation of various free parameters in our SED model, shown in Figure \ref{fig:jet_param_ephocs}. The temporal variation of B shown in the first panel of Figure \ref{fig:jet_param_ephocs} suggests a clear increase during the flaring state, mostly linked with the higher synchrotron emission. The second panel shows the variation of the bulk Lorentz factor of the jet ranges between 8 to 13 in different epochs, but no clear correlation is seen with the flux states. The temporal variation of the size of the emission region is shown in the third panel, where $R$ is larger in the quiet state compared to the flaring states, which suggests much faster variability in flaring states. As derived from SED modeling, the emission region is located outside the BLR but inside the DT for all the states. The energy densities of the disk, BLR, and DT are computed by \texttt{JetSeT} as described above, with the BLR following \citet{2003APh....18..377D}, and are given in the comoving frame at $R_H$. When the emission region is at a greater distance from the BLR region, a smaller number of BLR photons encounter the particle in the emitting region, and hence a lower gamma-ray flux can be seen. This can also be confirmed from panel 4, where we show the variation of BLR energy density, which varies depending upon the R$_H$ \citep{2009MNRAS.397..985G}. Since the emission region is located at a much farther distance from the SMBH or the accretion disk, and hence a very small number of accretion disk photons encounter the high-energy relativistic electrons and hence lower U$_{disk}$(panel 5 of Fig \ref{fig:jet_param_ephocs}), which results in a small contribution from the EC-disk. 
%If the emission region is produced closer to the base of the jet within the BLR during the flaring state, then the U$_{disk}$ is expected to be larger and as a result EC-disk contribution also rises, which is evident in  Fig~\ref{fig:sedfig}(b,c,d). 
The temporal variation of U$_{BLR}$ and U$_{Disk}$ is shown in the fourth and fifth panels of Figure \ref{fig:jet_param_ephocs}. It is evidently visible that the U$_{BLR}$, U$_{disk}$, and U$_{DT}$ have varied based on the emission region's geometry. The comoving energy densities depend on $\Gamma$ as well as on $R_H$ and $L_{disk}$. Inside $R_{DT}$ the DT energy density in the stationary frame is independent of $R_H$, $4.3\times10^{-5}$ erg cm$^{-3}$ for $L_{disk}=10^{46}$ erg s$^{-1}$, and the comoving value scales as $\Gamma^2$. The different $U_{DT}$ of F1, F2, and F3 therefore come only from their different $\Gamma$, with $U_{DT}/\Gamma^2 = 8.5\times10^{-5}$ erg cm$^{-3}$ in all three. In the stationary frame, $U_{Disk}$ scales with $L_{disk}/R_H^2$. The QS has a smaller $R_H$ than F3 but a smaller $U_{Disk}$, because the QS fit used a ten times lower $L_{disk}$. The same factor lowers $U_{DT}$ and $U_{BLR}$ of the QS.

 The total jet luminosity is calculated as a sum of L$_{\text{rad}}$, L$_{\text{kin}}$, and L$_{\text{B}}$, with the definitions given above; L$_{\text{e}}$ is the part of L$_{\text{kin}}$ carried by the relativistic electrons. We also compare the total jet luminosity for various flux states in the lower panel of Figure \ref{fig:jet_param_ephocs}. As expected, the luminosity is higher for the flaring states. The luminosity is minimum during the quiet state. To compare the energy of the jet with the Eddington luminosity, we first calculate the Eddington luminosity (L$_{Edd}$) using the formula 
\begin{equation}
    L_{Edd} = 1.26\times10^{38} \frac{M}{M_{\odot}} ~erg/s
\end{equation}
where M is the SMBH mass and M$_{\odot}$ is the solar mass. We found L$_{Edd} \approx$ $8.0 \times 10^{46}$ erg s$^{-1}$, taking $M$ = 10$^{8.8}$ M$_{\odot}$ \citep{2021ApJS..253...46P} and this is much higher than the total luminosity of the jet in different states evaluated from the broadband SED modeling. Our study examined the main cause of flaring events and found that the magnetic field, the Bulk Lorentz factor, the size of the emitting region, and the change in BLR and Disk energy densities, particle minimum and maximum energy, and particle spectral index are the main drivers. The variation of $B$ and $U_{BLR}$ follows exactly the same pattern, suggesting that the synchrotron and IC emission are highly correlated and show simultaneous flaring. 

Broadband SED modeling suggested that the synchrotron peak of flaring states lies between $10^{14}$ Hz and $10^{15}$ Hz with a jet luminosity around $10^{45}$ erg/s. It implies that 4C +27.50 is an Intermediate Synchrotron Peak (ISP) blazar, and it follows the standard blazar sequence as FSRQs generally have greater jet luminosity \citep{2001A&A...375..739D}. As \citet{2024ApJ...965..112R} found the source 4FGL J1520.8–0348 of almost similar redshift to be a peculiar type of blazar, and they concluded that it might be a blue FSRQ. In our source 4C +27.50, the synchrotron peak does not exceed $10^{15}$ Hz like 4FGL J1520.8–0348. In 4FGL J1520.8–0348, a state change from BL Lac to FSRQ has been observed, and both the synchrotron and IC peaks have the same flux. But in 4C +27.50, no state transition has been observed, and it showed a higher IC peak flux compared to the synchrotron peak flux. \citet{2019MNRAS.484L.104P} suggested that blue FSRQs have $\frac{L_{disk}}{L_{Edd}} > 0.01$. For 4C +27.50, we found $\frac{L_{disk}}{L_{Edd}} =$ 0.13, and it may tend towards the blue FSRQs class.

\begin{deluxetable*}{llllll}[ht]
%\digitalasset
\tablewidth{0pt}
\tablecaption{Broadband SED modeling best fit parameters. The Doppler factor is derived from $\Gamma$ and $\theta$. The jet powers are defined in Section~\ref{subsec:sed}. \label{tab:sed_params}}
\tablehead{
\colhead{\textbf{Symbol}} & \colhead{\textbf{Parameter [Units]}} & \colhead{\textbf{QS}} & \colhead{\textbf{F1}} & \colhead{\textbf{F2}} & \colhead{\textbf{F3}}
}
\startdata
\hline\hline
    \multicolumn{6}{c}{\textbf{Frozen parameters}} \\ \hline \hline
    $R_{BLR-in}$ & Inner radius of BLR [$10^{17}$ cm]  & 3.16 & 3.16 & 3.16 & 3.16 \\
    $R_{BLR-out}$ & Outer radius of BLR [$10^{17}$ cm]  & 3.48 & 3.48 & 3.48 & 3.48 \\
    $\tau_{\text{BLR}}$ & Optical depth of BLR & 0.1 & 0.1 & 0.1  & 0.1 \\
    $R_{DT}$ & Radius of DT [$10^{18}$ cm]  & 7.91 & 7.91 & 7.91  & 7.91 \\
    $T_{DT}$ & DT Temperature [$10^{3}$ K]  & 1.21 & 1.21 & 1.21 & 1.21 \\
    $\tau_{\text{DT}}$ & Optical depth of DT & 0.1 & 0.1 & 0.1 & 0.1 \\
    $L_{\text{Disk}}$ & Disk Luminosity [$10^{46}$ erg/s] & 0.1 & 1 & 1 & 1\\
    $T_{\text{Disk}}$ & Disk Temperature [$10^4$ K] & 3.97 & 3.97 & 3.97 & 3.97 \\ 
    $\theta$ & Jet viewing angle [degree] & 0.5 & 0.5 & 0.5 & 0.5 \\
    $z$ & Redshift & 1.2551 & 1.2551 & 1.2551 & 1.2551\\
    %$N_{\text{H\_cold\_to\_rel\_e}}$ & Cold proton to relativistic electron ratio & 0.1 & 0.1 & 0.1 & 0.1 \\ 
    \hline\hline
    \multicolumn{6}{c}{\textbf{Free parameters}} \\ \hline \hline
    $\gamma_{\min}$ & Low Energy Cut-Off & 270 & 110 & 130 & 92 \\
    $\gamma_{\max}$ & High Energy Cut-Off & $4.5\times10^{5}$ & $2.5\times10^{4}$ & $1.4\times10^{5}$ & $1.3\times10^{5}$ \\
    $N$ & Emitters Density [$10^2$/cm$^3$] & 0.33 & 4.4 & 16 & 5.5 \\
    $\alpha$ & Low-energy spectral index & 3.2 & 2.4 & 1.7 & 1.0 \\
    $\beta$ & Spectral curvature & 6.6 & 4.9 & 7.9 & 3.0 \\
    $\gamma_0$ & Turn-over Lorentz factor [$10^3$] & 4.9 & 2.5 & 4.6 & 1.4 \\
    $R$ & Emission Region Size [$10^{16}$ cm] & 6.2 & 1.4 & 0.79 & 1.1 \\
    $R_H$ & Emission Region Position [$10^{17}$cm] & 4.0 & 3.6 & 3.7 & 5.4 \\
    $B$ & Magnetic Field [G] & 0.39 & 1.2 & 0.87 & 0.51 \\
    $\Gamma$ & Jet Bulk Lorentz Factor & 11 & 12 & 8.5 & 13 \\
    $\delta$ & Doppler factor (derived) & 21 & 24 & 17 & 26 \\
    \hline\hline
    \multicolumn{6}{c}{\textbf{Energy Densities (comoving frame, at $R_H$)}} \\ \hline \hline
    $U_{\text{BLR}}$ & BLR Energy Density [$\text{erg/cm}^3$] & $3.2\times10^{-2}$ & 1.5 & 0.50 & $4.5\times10^{-2}$ \\
    $U_{\text{Disk}}$ & Disk Energy Density [$\text{erg/cm}^3$] & $1.9\times10^{-3}$ & $3.7\times10^{-2}$ & $2.2\times10^{-2}$ & $5.7\times10^{-3}$ \\
    $U_{\text{DT}}$ & DT Energy Density [$\text{erg/cm}^3$] & $9.9\times10^{-4}$ & $1.2\times10^{-2}$ & $6.1\times10^{-3}$ & $1.4\times10^{-2}$ \\
    $U_{\text{e}}$ & Electron Energy Density [$\text{erg/cm}^3$] & $1.3\times10^{-2}$ & 0.11 & 0.94 & 0.42 \\
    $U_{\text{B}}$ & Magnetic Field Energy Density [$\text{erg/cm}^3$] & $6.1\times10^{-3}$ & $5.5\times10^{-2}$ & $3.0\times10^{-2}$ & $1.0\times10^{-2}$ \\
    \hline \hline
    \multicolumn{6}{c}{\textbf{Jet Powers}} \\ \hline \hline
    $L_{\text{e}}$ & Jet power in relativistic electrons [$\text{erg/s}$] & $5.4\times10^{44}$ & $3.0\times10^{44}$ & $3.9\times10^{44}$ & $8.5\times10^{44}$ \\
    $L_{\text{rad}}$ & Jet radiative power [$\text{erg/s}$] & $2.9\times10^{44}$ & $9.2\times10^{44}$ & $1.9\times10^{45}$ & $1.1\times10^{45}$ \\
    $L_{\text{kin}}$ & Jet kinetic power (electrons and cold protons) [$\text{erg/s}$] & $7.5\times10^{44}$ & $4.8\times10^{44}$ & $4.9\times10^{44}$ & $1.0\times10^{45}$ \\
    $L_{\text{B}}$ & Jet power in magnetic field [$\text{erg/s}$] & $2.5\times10^{44}$ & $1.5\times10^{44}$ & $1.3\times10^{43}$ & $2.1\times10^{43}$ \\
    $L_{\text{Jet}}$ & Total jet power [$\text{erg/s}$] & $1.3\times10^{45}$ & $1.5\times10^{45}$ & $2.4\times10^{45}$ & $2.1\times10^{45}$ \\
\enddata
\end{deluxetable*}

\section{Summary} \label{sec:summary}
This is the first time the multi-wavelength light curve of 4C +27.50 has been studied. We selected the period from MJD 59215 to MJD 60751 because before 59215, the source was not detected in $\gamma$-rays. We identify four flaring states based on the BB algorithm, but the last flare was not followed in other wavebands due to the sun constraint. We selected three flaring episodes (namely F1, F2, and F3) and performed the multi-wavelength temporal and spectral studies. The archival optical and IR data from ZTF, ASAS-SN, and WISE are also collected to examine broadband behavior. A fractional variability study has been performed for each of the wavebands. ZTF r-band shows the highest fractional variability of 1.30 $\pm$ 0.01, followed by $\gamma$-ray fractional variability of 1.20 $\pm$ 0.01. Minimum flux doubling time was found to be 7.8 hours in the 1-day binned $\gamma$-ray light curve. This hour scale variability suggested the size of the emission region to be $\lesssim 8.9 \times 10^{15}$ cm and, in the colliding-shell scenario, its location to be $\gtrsim 1.1 \times 10^{17}$ cm away from the SMBH. The SED fits place the emission region just outside the BLR, at distances that correspond to the day-scale variability. A strong correlation between $\gamma$-ray and optical light curves has been found using ZDCF, suggesting a co-spatial origin of multi-wavelength emission. No strong correlation between $\gamma$-ray flux and index has been observed, but a high anti-correlation between X-ray flux and index has been found. Broadband SED modeling was performed for flaring and quiet states. The lower energy spectral index is found to be harder in flaring states than in the quiet state. The fitted parameters with respect to various flux states shown in Fig~\ref{fig:jet_param_ephocs} suggest that an increment in the magnetic field and electron energy density, with no systematic change in the bulk Lorentz factor, may have caused the flare in 4C +27.50. The total jet luminosity was found to be higher during the flaring state, but still lower than the Eddington luminosity of the source.

\begin{acknowledgments}
We thank the referee for their insightful and constructive comments on the manuscript. J.M. thanks Andrea Tramacere for the discussion about the JetSeT code. This research makes use of the publicly available data from Fermi-LAT obtained from the FSSC data server and distributed by NASA Goddard Space Flight Center (GSFC). This work made use of data supplied by the UK Swift Science Data Centre at the University of Leicester. The data, software, and web tools obtained from NASA’s High Energy Astrophysics Science Archive Research Center (HEASARC), a service of GSFC, are used in this work. This work made use of public data supplied by the NASA/IPAC Infrared Science Archive. This work made use of public optical data supplied by the ASAS-SN Sky Patrol Photometry Database. The authors acknowledge the support from the BHU IoE seed grant.
\end{acknowledgments}

\facilities{Fermi(LAT), Swift(XRT and UVOT), ASAS-SN, ZTF, NeoWISE}

\bibliography{sample7}{}
\bibliographystyle{aasjournalv7}

\end{document}